\documentclass[pra,aps,showpacs,groupauthors,footinbib,twocolumn]{revtex4-1}

\usepackage{amsmath,amsfonts,amssymb,amsthm}
\usepackage{graphicx}

\usepackage{amssymb}
\usepackage[english]{babel}
\usepackage{amsfonts}
\usepackage{amsmath}
\usepackage{graphicx}
\usepackage{bm}
\usepackage{color}

\newcommand{\ket}[1]{|{#1}\rangle}

\newcommand{\tunnu}[1]{\textcolor{black}{#1}}

\usepackage{ulem}

\begin{document}

%\preprint{APS/123-QED}

%%

%%---- Title of the paper --------------------------------------------------------------------------------

%%

\title{Speeding up the spatial adiabatic passage of matter waves in optical microtraps by optimal control}

%%

%%---- Authors and affiliations --------------------------------------------------------------------------

%%

\author{Antonio Negretti$^1$}
\author{Albert Benseny$^2$}
\author{Jordi Mompart$^2$}
\author{Tommaso Calarco$^1$}
\affiliation{$^1$Institut f\"ur Quanteninformationsverarbeitung, Universit\"at Ulm, D-89069 Ulm, Germany}
\affiliation{$^{2}$Grup d'\`{O}ptica, Departament de F\'{i}sica, Universitat Aut\`{o}noma de Barcelona, E-08193 Bellaterra, Spain}
\date{\today}

%%

%%---- Abstract ------------------------------------------------------------------------------------------

%%

\begin{abstract}
We numerically investigate the performance of atomic transport in optical microtraps via the so called spatial adiabatic 
passage technique. Our analysis is carried out by means of optimal control methods, which enable us to determine suitable 
transport control pulses. We investigate the ultimate limits of the optimal control in speeding up the transport process 
in a triple well configuration for both a single atomic wave packet and a Bose-Einstein condensate within a regime of 
experimental parameters achievable with current optical technology. 
\end{abstract}

%%

%%---- PACS numbers --------------------------------------------------------------------------------------

%%


%% 03.67.Lx : Quantum computation architectures and implementations
 
\maketitle

%%

%%---- Section --------------------------------------------------------------------------------------

%%

\section{Introduction}

The coherent control of matter waves has become a very relevant research topic with significant technological applications 
such as in atom lasers~\cite{Bloch1999,Gustavson2001,Haensel2001} and quantum information processing 
(QIP)~\cite{QIP:ACbook11}, to only cite a few. Indeed, very recently, experimental demonstrations with 
ultracold atoms in two-dimensional (2D) optical lattices, of single quantum bit (qubit) rotations~\cite{QIP:Lundblad09}, 
single site addressability~\cite{Bakr2010,Weitenberg2011}, and single atom 
detection~\cite{QIP:Sherson10,QIP:Bakr09}, with very high fidelities, have been reported.
One of the current most important goals of QIP is to go beyond the manipulation of a handful of qubits 
since the main challenge is to build scalable quantum hardware where several thousands of qubits 
are coherently manipulated within the relaxation times of the system~\cite{QIP:Divincenzo00}. 
It has been also understood that the practical realization of QIP requires to devise new architectures. Indeed, 
several schemes for two-qubit quantum gates implementations based on atomic systems proposed in 
the past can be, at least in principle, realized in experiments, but they may present several shortcomings 
when building a scalable quantum computing hardware (see Ref.~\cite{Negretti2011} for a review on QIP 
with neutral particles). Moreover, in order to name a QIP system ``scalable'', it is also important that the resources 
required to control the quantum system, typically classical devices (e.g., laser fields, refrigerators), are 
scalable as well~\cite{Ladd2010}. 

Paradigmatic examples of QIP are the Cirac-Zoller~\cite{Cirac1995}  and the M\o lmer-S\o rensen~\cite{Molmer1999} 
ion quantum computers, which have been proven to be powerful schemes to experimentally realize small 
quantum algorithms~\cite{Gulde2003,Kaler2003,Sackett2000} or to engineer quite exotic entangled states 
(up to 14 qubits), like Werner ~\cite{Haeffner2005} or Greenberger-Horne-Zeilinger states~\cite{Leibfried2005,Monz2011}. 
These schemes, however, present rather difficult technical problems when hundreds or even thousands of ions participate 
in the collective motion (e.g, decoherence of motional modes, sensitivity to electric noise~\cite{Hughes1996}). 
Thus, currently, a big effort is made in the design and practical realization of new schemes that are actually scalable. 
For instance, in Ref.~\cite{Kielpinski2002} it has been proposed an ion quantum processor architecture, where some 
areas of the chip processor are used only to store the information, and others to manipulate it. Such a design 
requires to transport an ion from one location to another one in the chip preserving its quantum mechanical 
coherence. A similar problem is encountered for QIP implementations with neutral atoms either in optical 
lattices~\cite{QIP:Jaksch99,QIP:Charron02,QIP:Treutlein06a} or in microwave atom 
chips~\cite{QIP:Treutlein06b,QIP:Boehi09}. Atoms trapped in optical lattices can be efficiently prepared via 
the superfluid Mott insulator quantum phase transition~\cite{QIP:Jaksch98,Greiner2002}, and single sites 
addressed~\cite{Bakr2010,Weitenberg2011}, but the realization of quantum gates between qubits located 
in far away lattice sites can be a serious problem for a scalable neutral-atom-based quantum processor. Solutions 
to this issue can be afforded, for instance, by auxiliary atoms that can be efficiently transported in state-independent 
periodic external traps~\cite{QIP:Calarco04} or by using optical tweezers~\cite{Weitenberg2011b}.

A major underlying concept of these QIP paradigms is the coherent transport of ions or atoms in such a way that they 
mediate the operation of quantum gates between spatially distant qubits. The needed transport time, however, has been 
estimated to be about 95\% of the time used for carrying out the whole quantum computation~\cite{Huber2008}. 
It is therefore imperative to reduce the time needed to transport an atom or ion from the quantum memory to the 
processing units and, therefore, to engineer robust control transport pulses. In this respect, optimal control theory is a 
prominent candidate for a drastic improvement of the design of accurate QIP protocols, and, recently, several 
theoretical investigations on the optimal transport of both a single atom and an atomic ensemble have been 
undertaken~\cite{Hohenester2007,Murphy2009,Chen2011,Torrontegui2011}. Besides this, very recently, control 
pulses numerically obtained by using iterative optimization algorithms have been experimentally applied, with great 
success, in order to efficiently transfer a one-dimensional (1D) degenerate Bose gas from the transverse ground 
to the lower excited state of a waveguide potential~\cite{Buecker2011}. This result shows the potential offered 
by optimal control methods to engineer current experiments of ultra-cold atoms.

\begin{figure}[t]
\begin{center}
\includegraphics{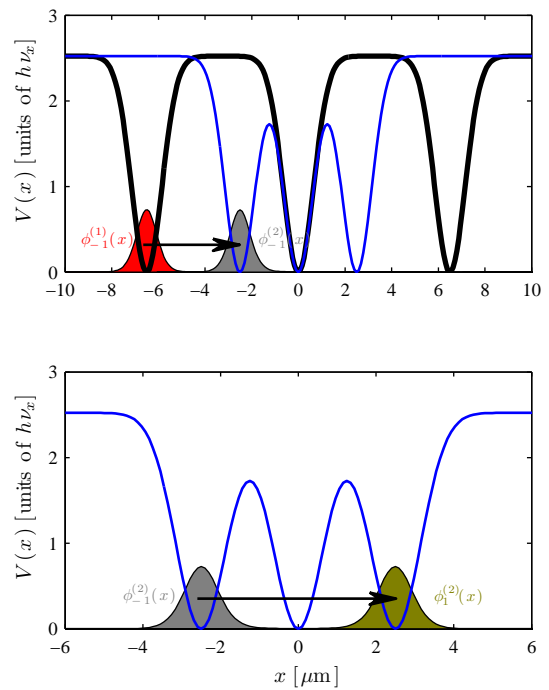}
\end{center}
\caption{(Color online). Upper panel: Initial (black thick line) and final (blue thin line) potential configuration for 
step 1 of the transport process with trap separation 6.5 $\mu$m, beam waist $2 w = 1.3\,\mu$m, potential depth 
$V_0 = k_B\times 86\,\mathrm{nK}$, and trap frequency $\omega_x = \sqrt{4V_0/m\sigma^2} = 2\pi\nu_x \simeq 2 \pi\times 711$ Hz. 
The initial ground state wave function $\psi_0(x)\equiv\phi^{(1)}_{-1}(x)$ (red) of the left well is also displayed.
Lower panel: Initial and final potential configuration (same as the blue thin line of the upper panel) for the SAP process. 
The goal wave function $\psi_{\mathrm{g}}^{\prime}(x)\equiv\phi^{(2)}_{-1}(x)$ (bright grey) of step 1 is shown. 
This state also corresponds to the initial wave function of the SAP process, whose goal wave function is 
$\psi_{\mathrm{g}}(x)\equiv\phi^{(2)}_1(x)$ (dark green), superimposed on the right well of the lower 
panel. The horizontal black arrows indicate the transport direction.}
\label{fig:pot}
\end{figure}

In this work we investigate the ultimate limits of the transport of neutral atoms in optical microtrap arrays by means of 
numerical optimization methods. Specifically, we are interested in the spatial adiabatic passage (SAP) protocol \cite{Eckert2004}, 
the matter wave analogue of the stimulated Raman adiabatic passage (STIRAP) technique used in quantum optics to transfer the 
population between two atomic internal levels~\cite{Bergmann1998}. The SAP technique consists in adiabatically following an 
energy eigenstate of the system, the so-called spatial dark state, that only involves the vibrational ground states of the two extreme 
wells of a triple-well potential (see Fig.~\ref{fig:pot}). The spatial dark state presents at all times a node in the central region such 
that the middle-well population is almost negligible throughout the transport process. 
Therefore, the SAP protocol enables to transport an atom from a lattice site to the next-nearest-neighbor without populating the 
nearest-neighbor site (the middle well in Fig.~\ref{fig:pot}). If one atom is present in the middle trap, the SAP technique can be used 
to implement a single atom diode or a transistor \cite{Benseny2010}.
In addition, this transport technique may allow to reduce the complexity of several quantum computing 
architectures~\cite{QIP:Mompart03,QIP:Treutlein06b,QIP:Charron06,Weitenberg2011,QIP:Brennen99}. In fact, 
compared to the tunneling induced oscillation between two adjacent traps, such technique is more robust against 
variations of the system parameters and requires less precise control of the distance and timing. 
Three-level atom optics techniques \cite{Eckert2004}, such as the SAP protocol, allow also to create superpositions 
(spatial dark states) of matter waves between two separated sites of the lattice, useful for applications in atomic interferometry, or to 
inhibit the tunneling among lattice sites, therefore allowing 
to create conditional phase shifts for quantum logic, or to transport an empty site~\cite{Eckert2004}. 
We also mention, that, recently, the implementation of the SAP protocol for radio-frequency traps~\cite{Zobay2001,Lesanovsky2006,Hofferberth2007} 
within the three mode approximation has been investigated~\cite{Morgan2011}, and that such technique could be 
even employed for the ion transport in segmented microtraps~\cite{Schulz2006,Reichle2006}.

Beside this, we will investigate the transport of a Bose-Einstein condensate (BEC) in relation to recent experiments 
with optical dipole traps~\cite{QIP:Lengwenus07,Lengwenus2010,Kruse2010}. We note that a similar study has been carried out in 
Ref.~\cite{Hohenester2007}, where the optimal transport of a BEC in magnetic microtraps, like the ones produced 
with atom chips~\cite{QIP:ACbook11}, has been investigated, and that, very recently, the optimal control pulses 
for harmonically trapped BECs have been analytically determined~\cite{Torrontegui2011}. We underscore that, while the 
goal of those investigations was to transfer a BEC between spatially separated locations, here, in addition to this goal, 
we aim at minimizing the occupancy of the middle well in a triple-well configuration, as showed in Fig.~\ref{fig:pot}, by 
following as much as possible the spatial dark state of the trapping potential. This additional constraint is the main signature 
of the SAP protocol.

%%

%%---- Section --------------------------------------------------------------------------------------

%%

\section{Optical dipole microtraps}
\label{sec:dipoletrap}

We consider a (transverse) potential, where either a single atom or BEC is trapped, given by the following analytical expression 

\begin{equation}
V(x,t) = V_0\left\{1 - \sum_{k=-1}^1 v_k(t)\exp\left[
-\frac{(x-k d_k(t))^2}{2 w^2}
\right]
\right\},
\label{eq:Vxt}
\end{equation} 
where $V_0$ represents the depth of the three Gaussian dipole traps, with $2 w$ being the laser beam waist, 
$d_{-1}(t)$ represents the distance between the central trap and the left trap, $d_1(t)$ is the distance between 
the central trap and the right trap, while the central trap remains at $d_0(t)\equiv 0$ $\forall t\in [0,T]$.
Here $T$ is the time needed to transport the system of interest (i.e., an atom or a 
BEC) initially prepared in the ground state of the trap on the left (centered in $-\vert x_0^{\prime}\vert$) 
to the ground state of the trap on the right (centered in $\vert x_0^{\prime}\vert$).
As outlined above, we shall first consider a 1D scenario, but later we shall also study the influence 
of the dimensionality on the transport performance. 
\tunnu{We note that optical dipole traps, as the ones of Ref.~\cite{Lengwenus2010}, can be designed to form a 
2D lattice. The $x-y$ plane, that defines the lattice, has a weaker confinement than the 
(vertical) axial direction, thereby defining a ``pancake" geometry. It is in 
the (transverse) $x-y$ plane that the transport occurs and because of this we refer to the potential (\ref{eq:Vxt}) 
along $x$ as transverse (in our study the trap frequencies in the $x$ and $y$ directions will be assumed to be equal).}

In Fig.~\ref{fig:pot} the potential is illustrated, where experimentally realistic parameters \tunnu{(i.e., potential 
depth, trap separation, and beam waist)} have been considered for $^{85}$Rb atoms.
Due to the large (initial) separation between the trap minima of the lattice [$x_0^{\prime}=6.5\,\mu$m, see 
Fig.~\ref{fig:pot} (top)], no tunneling is expected to happen until the traps are closer. The SAP transport is 
split and optimized in three different stages. Firstly, in a time $T_1$, the initial atomic state (the ground state 
of the left well) is moved from $x_0^{\prime}=-6.5\,\mu$m to $x_0=-2.5\,\mu$m, i.e., from the left well of the 
initial trap configuration (thick black line) given 
in Fig.~\ref{fig:pot} (top) to the left well of the potential of the upper panel of Fig.~\ref{fig:pot} (blue thin line).
Secondly, the atomic system is brought, in a time $T_2$, from the left well centered at $x_0=-2.5\,\mu$m to 
the right one centered at $x_0=2.5\,\mu$m (see blue line of the lower panel of Fig.~\ref{fig:pot}).
The third step consists 
in bringing the system from the right well centered at $x_0= 2.5\,\mu$m to the right well centered at 
$x_0^{\prime}=-6.5\,\mu$m of the potential displayed in Fig.~\ref{fig:pot} (top), which is equivalent to 
reversing the first step of the process.
Hence, the total transport time is $T=2T_1+T_2$.
It is in the second step that the SAP process takes place.
Since at the beginning the atoms are quite far apart, the direct application of the SAP technique would simply 
slow down the whole transport process, because initially no tunneling would take place.
Instead, with the above outlined procedure, the first step can be sped up as much as possible until the 
so-called quantum speed limit is reached, that is, the minimum time 
required to evolve a quantum system from an initial state to an other orthogonal 
state~\cite{Bhattacharyya1983,Giovannetti2003,Caneva2009}.

\begin{figure}[t]
\begin{center}
\includegraphics{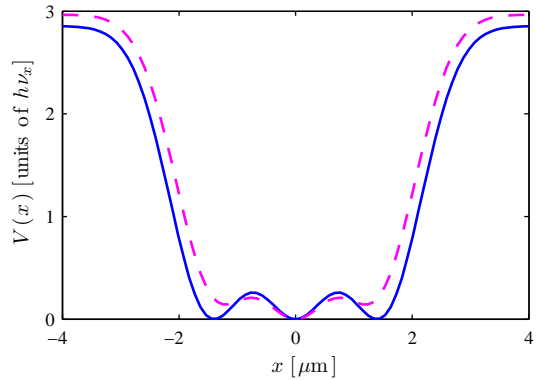}
\end{center}
\caption{(Color online). Potentials at time $t=T_2/2$: the blue (solid) line shows the potential when $v_{\pm 1}$ 
are time-dependent, whereas the magenta (dashed) line when $v_{\pm 1}$ are time-independent. The potential 
minima have been lifted by about $0.33 h\nu_x$ and $0.44 h\nu_x$, respectively. The 
symmetry of the potentials is due to the fact that, at that time, the two outer traps are equidistant from the 
centre $x=0$, that is, a distance almost equal to the minimal allowed trap separation $\delta x_0 = 1.43\,\mu$m. 
The rest of trap parameters are as in Fig.~\ref{fig:pot}.}
\label{fig:potT2}
\end{figure}

Finally, we note that a crucial condition for the realization of SAP is that the initial and final states involved during the transport 
process should be in resonance. This requirement can be fulfilled by fixing at all times the minima of the potential at the 
same energy level, through the control of the time-dependent parameters $v_{-1}(t)$, $v_{1}(t)$ [$v_0(t)\equiv 1$] 
(see also Fig.~\ref{fig:potT2}), as well as by fixing the maxima of the triple well configuration to the same energy 
level, through the control of $w$, that is, the beam waist (see the appendix for the analytical expressions of 
$v_{\pm 1}$). The control of the latter, however, would require beam waist values below the 
actual experimentally achievable limit~\cite{Lengwenus2010}, and therefore it will not be considered in our study. 
Hence, in our analysis we have fixed $w$ to the minimum experimentally achievable value \tunnu{(i.e., 0.65 $\mu$m)}, 
which enables us to prepare the atoms, before the transport, in a lattice configuration with minimum periodicity.

%%

%%---- Section --------------------------------------------------------------------------------------

%%

\section{Single atom transport}

In this section we analyze the transport of a single trapped atom in the absence of a thermal or quantum bath. 
The atomic state obeys the Schr\"odinger equation of motion.
In order to speed up the transport process we rely on numerical optimization techniques. For the problem considered 
in this paper, we employed a recently 
introduced optimization method, named the chopped random basis (CRAB) algorithm~\cite{Caneva2011}. Such a 
method has been shown to be a powerful tool in order to optimize the closed dynamics of many-body quantum 
systems~\cite{Doria2011} and the dynamics of light harvesting~\cite{Caruso2011}. Moreover, since the implementation 
of the CRAB algorithm does not rely on the equation of motion that governs the system dynamics, there is no need of 
algorithmic modifications when nonlinear dynamics is regarded (e.g., the dynamics of a BEC), unlike for the monotonically 
convergent Krotov algorithm~\cite{Krotov1996,Sklarz2002,Reich2010} or the gradient ascent pulse engineering 
algorithm~\cite{Khaneja2005}. Even though CRAB does not provide monotonic convergence it allows to directly restrict 
and select the space of control pulses (e.g., enforcing limited bandwidth), since it relies on a multi-variable function minimization that 
can be performed, for example, via a direct-search method (e.g., the Nelder--Mead method as implemented, for instance, in 
MATLAB). For more details on the procedure of the CRAB algorithm implementation and its computational performance 
we refer to Ref.~\cite{Caneva2011}.

\subsection{Optimization of step 1 of the transport process}
\label{sec:sp1}

We remind that our goal here is to transport an atom initially prepared in the ground state of the left well of the potential 
displayed in Fig.~\ref{fig:pot} (top), centered in $x_0^{\prime}=-6.5\,\mu$m, to the ground state of the left well of the 
potential shown in Fig.~\ref{fig:pot} (bottom) blue (solid) line, centered in $x_0=-2.5\,\mu$m. In this case we shall consider 
the control pulses to be identical, that is, $d_{-1}(t)=d_1(t)\equiv d(t)$. 

In this section, the objective functional (to be minimized) for the control problem we are interested in can be identified by the 
so-called overlap infidelity, namely
\begin{eqnarray}
\mathcal{I}=1-\left\vert
\int_{\mathbb{R}}\mathrm{d}x\,\psi_{\mathrm g}^{\prime *}(x)\psi(x,T_1)
\right\vert^2.
\label{eq:infidelity}
\end{eqnarray}
Here $\psi(x,T_1)$ is the wave function at time $t=T_1$  propagated from the initial condition $\psi_0(x)\equiv\phi_{-1}^{(1)}(x)$ 
at time $t=0$, where $\phi_{-1}^{(1)}(x)$ is the ground state of the left trap centered at $x=x_0^{\prime}=-6.5\,\mu$m. The wave 
function $\psi_{\mathrm g}^{\prime}(x)$ is the wave function we aim to achieve in a given time $T_1$, that is, the ground state 
$\phi_{-1}^{(2)}(x)$ of the left well centered at $x=x_0=-2.5\,\mu$m.
\tunnu{The superscript $(j)$ in $\phi_{-1}^{(j)}$ refers to the two first stages of the transport process. When $j=1$ the state 
$\phi_{-1}^{(1)}$ corresponds to the ground state of the left well of the potential (thick black line) illustrated in the upper panel 
of Fig.~\ref{fig:pot}, while for $j=2$ it corresponds to the ground state of the left well of the potential shown in the lower panel 
of Fig.~\ref{fig:pot} (blue solid line). The same applies to $\phi_1^{(j)}$, but for the right wells.}

\begin{figure}[t]
\begin{center}
\includegraphics{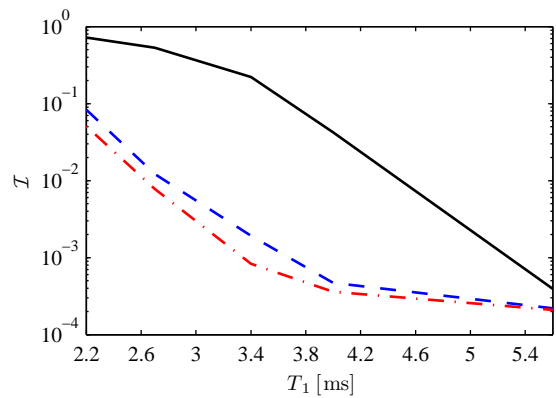}
\end{center}
\caption{(Color online).  Overlap infidelity vs. transport time: black (solid) line with the control pulse $d_{\mathsf{ho}}(t)$ 
defined in Eq.~(\ref{eq:michael}); the other two lines show the infidelity obtained with the CRAB optimized control pulse 
$d_{\mathsf{opt}}(t)=d_{\mathsf{ho}}(t)g_{\mathsf{opt}}(t)$ for $N_g=8$ (dashed line) and $N_g=16$ (dot-dashed line).}
\label{fig:tab1}
\end{figure}

The wells of  the upper panel of Fig.~\ref{fig:pot} (thick and thin lines) are sufficiently deep that the trapping potentials 
can be, with good  approximation, considered harmonic, but with slightly different trap frequencies.
Hence, an excellent guess control pulse is given by~\cite{Murphy2009}
\begin{eqnarray}
d_{\mathsf{ho}}(t)&=&(x_0^{\prime}-x_0)\left\{
\frac{t}{T_1}+\sin\left(\frac{2\pi t}{T_1}\right)\left[\frac{8\pi}{3(\omega_xT_1)^2}-\frac{2}{3\pi}\right]\right.\nonumber\\
&-&\left.\sin\left(\frac{4\pi t}{T_1}\right)\left[\frac{4\pi}{3(\omega_xT_1)^2}-\frac{1}{12\pi}\right]
\right\} -x_0^{\prime}.
\label{eq:michael}
\end{eqnarray}
For a particle in a moving harmonic potential such a control pulse is optimal, i.e., it yields $\mathcal{I}=0$, and it is 
quite robust against control pulse distortions~\cite{Murphy2009}.
Nevertheless, as Fig.~\ref{fig:tab1} shows (solid line), for our case with Gaussian traps, we obtain already 
a good result for large, but  not adiabatic (i.e., not in the regime $\omega_xT_1\gg 1$), transport times. 

To reduce the infidelity we further optimized the transport process by means of the CRAB algorithm, which works as follows: we start with the 
initial guess given by Eq.~(\ref{eq:michael}) and we define the new control pulse as $d(t)=d_{\mathsf{ho}}(t)g(t)$, where
\begin{eqnarray}
\label{eq:gt}
g(t) = 1+\frac{1}{\lambda(t)}\sum_{k=1}^{N_g}\left[A_k\sin\left(\omega_k t\right)+B_k\cos\left(\omega_kt\right)\right].
\end{eqnarray}
Here $\omega_k=2\pi k/T_1$, $N_g\in\mathbb{N}$ is the number of time-independent $A_k$ and $B_k$ coefficients, $\lambda(t)$ is a 
time-dependent function enforcing the boundary conditions of $d(t)$ at $t=0$ and $t=T_1$, namely $\lim_{t\rightarrow 0,T_1}\lambda(t)=+\infty$. 
Basically, the CRAB algorithm seeks for the time-independent coefficients $A_k$, $B_k$ and frequencies $\omega_k$ that minimize the 
overlap infidelity~(\ref{eq:infidelity}). 
Besides, in the numerical simulations, we set a tolerance ($\sim 10^{-4}$) on the determination of either the coefficients or the frequencies 
$\omega_k$. Such a tolerance is defined as the minimum allowed distance between the vertexes of the polytope generated within the Nelder-Mead 
multidimensional non-linear minimization procedure~\cite{Press2007}.

As illustrated in Fig.~\ref{fig:tab1}, the CRAB algorithm slightly improves the result for large $T_1$ times obtained with the control pulse defined 
in Eq.~(\ref{eq:michael}), but for short times the slight difference due to the proximity of the central trap in the trap frequencies 
(of about 0.7\%) of 
the left wells [thick and thin lines of Fig.~\ref{fig:pot} (top)], becomes crucial as well as the anharmonicity of the confinement potential. 
We have also investigated the improvement of the overlap infidelity due to a higher number of harmonics $N_g$ involved in the control pulse. As 
Fig.~\ref{fig:tab1} shows, 
a significant improvement can be observed only for initial large overlap infidelities, whereas for already almost perfect transport processes the 
reduction is almost insignificant. This numerical observation is not surprising, since the dynamics with an initial large overlap infidelity 
require more sophisticated control pulses (i.e., higher harmonics), in order to properly steer the atomic dynamics. Such more complex 
control pulses, however, might be more difficult to implement experimentally. Beside this, as shown in Fig.~\ref{fig:tab1}, 
we also note that the overlap infidelity drops quite significantly, as a rule of thumb, for times larger than 3 ms, which is consistent with the fact 
that the transport time cannot be shorter than the inverse of the typical trap frequency, that is, $1/\nu_x\sim 2$ ms. We underscore that the 
threshold $\nu_x^{-1}$ cannot be precisely identified with the quantum speed limit, but it is in close relation with it. An exact determination of the 
quantum speed limit relies on the time average of the instantaneous energy fluctuations, which depend on the particular control pulse. This is 
non trivial computational task for time-dependent Hamiltonians. Only in simple cases, such as the Landau-Zener model, the 
minimum time can be efficiently estimated~\cite{Caneva2009}. However, further optimization, that is, larger values of $N_g$, will not overcome this 
(physical) limit.

\begin{figure}[t]
\begin{center}
\includegraphics{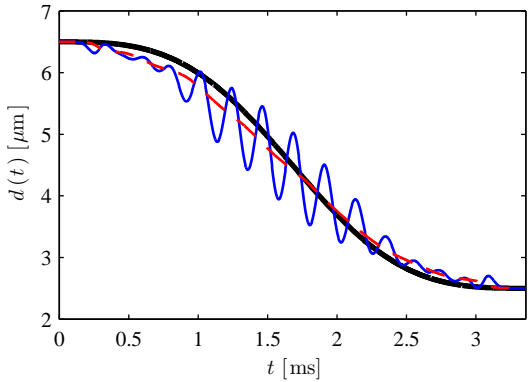}
\end{center}
\caption{(Color online). Optimal control pulses of step 1 of the transport process (red dashed line with $N_g=8$, blue solid line $N_g=16$); 
initial guess given by Eq.~(\ref{eq:michael}) (black thick solid line). The transport time is $T_1=3.4$ ms.}
\label{fig:optd}
\end{figure}

As an example of the optimization, Fig.~\ref{fig:optd} shows the optimized transport control pulses for two different sets of coefficients 
$\{A_k,B_k\}$ for the transport time $T_1=3.4$ ms, where $d_{\mathsf{ho}}(t)$ has been used as initial guess (black thick line). We see that, to achieve very small infidelities, the control pulse involves more wiggles, especially of large amplitude at the intermediate times, where the 
system is more excited. On the other hand, at the end of the transport process the system has to be restored again in the ground state 
of the trap, and therefore the modulation of the control pulse is more ``gentle".

\begin{figure}[t]
\begin{center}
\includegraphics{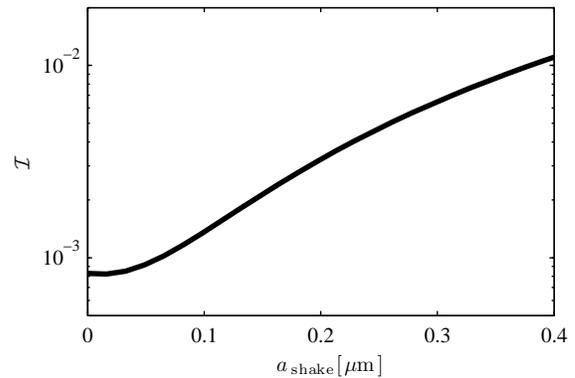}
\end{center}
\caption{(Color online). Overlap infidelity vs. amplitude of a shaking in the trap positions of the outer traps with 
$\omega_{\mathrm{shake}}=10^{-2}\omega_x$, $\delta x_0=1.43\,\mu$m, and $\ell=\sqrt{\hbar/m\omega_x}=0.41\,\mu$m. 
The transport time is $T_1=3.4$ ms and $N_g=16$.}
\label{fig:robust}
\end{figure}

We have also investigated the robustness against trap position fluctuations due to possible experimental imperfections. 
To this aim we used for the time-dependent control pulses the following expression
\begin{eqnarray}
d_{\mp 1}(t)&=&d_{\mathsf{opt}}(t)\pm a_{\mathrm{shake}}\sin(\omega_{\mathrm{shake}}t),
\label{eq:shake}
\end{eqnarray}
where $d_{\mathsf{opt}}(t)$ is the optimal control pulse obtained with CRAB. With such a choice, for $a_{\mathrm{shake}}>0$ the left 
and right wells oscillate in phase, whereas for $a_{\mathrm{shake}}<0$ their oscillations are out of phase. 
However, since in this particular step of the transport process only a populated well is effectively involved, the only value that matters is 
$\vert a_{\mathrm{shake}}\vert$ (for negative values the behavior of $\mathcal{I}$ is basically the same). The result of such analysis is 
shown in Fig.~\ref{fig:robust} for $\omega_{\mathrm{shake}}/\omega_x=10^{-2}$ and $T_1=3.4$ ms (see also Fig.~\ref{fig:tab1}). 
We see that the optimal control pulse is quite robust against fluctuations of the trap position. This result is in agreement with the findings 
for a particle in a moving harmonic potential~\cite{Murphy2009}.

Finally, we investigated the role of spatial dimensionality. Up to now, we performed our analyses in the quasi-1D regime. 
\tunnu{However, we recall that in the experiments of Refs.~\cite{QIP:Lengwenus07,Lengwenus2010} the potential in the $z$ 
(axial) direction is shallower than in $x$ or $y$.} We therefore performed numerical simulations of the 2D Schr\"odinger 
equation with the trapping potential
\begin{eqnarray}
V(x,z,t) = V_0\left\{1 - \sum_{k=-1}^1 v_k(t)e^{
-\frac{[x-k d(t)]^2}{2 w^2}}e^{-\frac{z^2}{2w_{z}^2}}
\right\}.
\label{eq:Vxyt}
\end{eqnarray} 
Here $2 w_{z}$ is the beam waist along the $z$ direction. The ratio $\omega_{z}/\omega_x$ is determined by $w_x/w_{z}$, 
namely, the larger $w_{z}$ is, the smaller the (axial) frequency $\omega_{z}$. The new (ground) initial and goal states have been 
obtained by using the imaginary time propagation procedure, typically used for the determination of the ground state of a BEC.
As trial functions for the imaginary time propagation we took the tensor product of the solutions of the quasi-1D regime: for the transverse 
direction by a numerical exact diagonalization of the single particle Hamiltonian, and for the axial direction by choosing the Gaussian 
ground state of the harmonic oscillator.  

\begin{figure}[t]
\begin{center}
\includegraphics{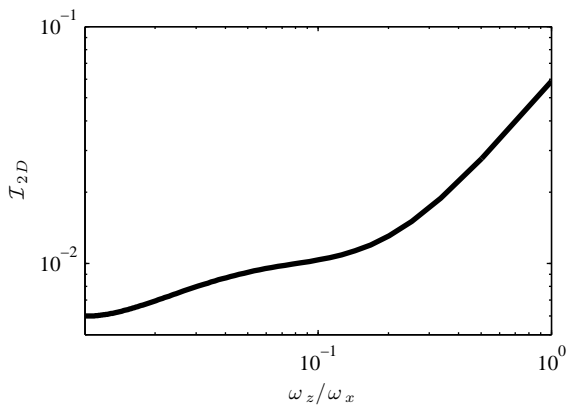}
\end{center}
\caption{(Color online). Overlap infidelity vs. the ratio $\omega_{z}/\omega_x$. Here $\mathcal{I}_{2D}$ is simply the 
generalization of Eq.~(\ref{eq:infidelity}) with a two-variable integration. The transport time is $T_1=3.4$ ms and $N_g=16$.}
\label{fig:check2D}
\end{figure}

The result of such analysis, for the optimal control pulse obtained for the transport time $T_1=3.4$ ms with $N_g=16$, is shown in 
Fig.~\ref{fig:check2D}. As it is illustrated, the smaller $\omega_{z}$ is, the better the infidelity. This behavior is reasonable: since 
$w_{z}\gg w_x$, we can write $\exp(-z^2/2w_{z}^2)\simeq 1-z^2/2w_{z}^2$, which implies an almost perfect 
harmonic potential in the axial direction, thereby almost separable from the one in the transverse direction. The figure shows, however, 
that the infidelity becomes larger than the one obtained in the quasi-1D regime, which is almost one order of magnitude smaller (see 
Fig.~\ref{fig:tab1}).  We attribute this enhancement to the not completely negligible coupling between the axial and transverse motion. Thus, for 
an exact infidelity and control pulse evaluation, a 2D optimization should be performed. Of course, the result relies on the particular chosen 
$\omega_{z}/\omega_x$ ratio, which ultimately is determined by the experimental setup. As already pointed out in 
Ref.~\cite{QIP:Treutlein06b}, care has to be taken when calculations with realistic parameters in the quasi-1D regime are considered. 
The quantum speed limit behavior, however, remains fundamentally the same as the one of the quasi-1D regime outlined above, and 
the control pulses obtained in this regime would be excellent initial guesses for the 2D optimization. 

\subsection{Optimization of step 2: SAP process}

The transport via SAP is the second step of the optimization process outlined above, where  the initial condition at time $t=0$ 
is given by $\psi_0(x)\equiv\phi_{-1}^{(2)}(x)$, whereas $\psi_{\mathrm g}(x)$ is the ground state wave function $\phi_1^{(2)}(x)$ of 
the right well centered at $x=x_0$ (see Fig.~\ref{fig:pot}). 

We start the optimization by using the following initial guess control pulses: 
\begin{eqnarray}
d_1^0(t) = \left\{
\begin{array}{l}
(x_0-\delta x_0)\cos^{2}\left(\frac{\pi t}{T_2-t_{\mathrm{d}}}\right)+
\delta x_0 
\hspace{0.3cm} t\in[0,T_2-t_{\mathrm{d}}]\\
x_0
\hspace{4.3cm}t\in(T_2-t_{\mathrm{d}},T_2]
\end{array}
\right.\nonumber\\
\end{eqnarray}
\begin{eqnarray}
d_{-1}^0(t) = \left\{
\begin{array}{l}
x_0
\hspace{4.95cm}t\in[0,t_{\mathrm{d}}]\\
(x_0-\delta x_0)\cos^{2}\left(\frac{\pi (T_2-t)}{T_2-t_{\mathrm{d}}}\right)+
\delta x_0 
\hspace{0.5cm} t\in(t_{\mathrm{d}},T_2].
\end{array}
\right.\nonumber\\
\end{eqnarray}
We note that $d_1(t)$ is $d_{-1}(t)$ time inverted. We chose the time delay between the two control pulses as 
$t_{\mathrm{d}}= 0.17\,T_2$, where $T_2$ is the transport time used to carry out the SAP technique. Such a choice 
is due to the analogy between SAP and STIRAP. Indeed, as shown in Ref.~\cite{Bergmann1998}, the time 
over which the two control pulses do overlap, has to fulfill the (adiabatic) criterion

\begin{eqnarray}
t_{\mathrm{d}}>\frac{10}{\min_{t\in [0,T]}\{\Omega(d(t)/\ell_x)\}},
\label{eq:td}
\end{eqnarray}
where~\cite{Eckert2004}

\begin{equation}
\frac{\Omega(d/\ell_x)}{\omega_x}=2\frac{d}{\ell_x}\left(\frac{e^{(d/\ell_x)^2}\{1+d[1-\mathrm{erf}(d/\ell_x)]/\ell_x\}-1}
{\sqrt{\pi}[e^{2(d/\ell_x)^2}-1]}\right),
\label{eq:Rabi}
\end{equation}
with $\ell_x=\sqrt{\hbar/m\omega_x}$. The tunneling ``Rabi'' frequency describes the coupling between 
the left and the middle wells for $d=d_{-1}^0$ and between the right and the middle wells for $d=d_1^0$. 
We note, however, that Eq.~(\ref{eq:Rabi}) is only valid for harmonic trapping potentials. For Gaussian traps 
the actual Rabi frequency must be numerically assessed, but for an estimation of the time delay, Eq.~(\ref{eq:Rabi}) 
can be used.
As we will discuss at the end of this section about the robustness of SAP against fluctuations 
of $t_{\mathrm{d}}$, the error induced by using Eqs.~(\ref{eq:td},\ref{eq:Rabi}) is indeed small, and the value of 
$t_{\mathrm{d}}$ used in our analyses is quite reasonable.

In this scenario the CRAB optimization works as follows. For the guess control pulse $d_{-1}^0(t)$ we take
\begin{eqnarray}
\label{eq:dpmguess}
d_{-1}(t) = \left\{
\begin{array}{l}
x_0
\hspace{5.20cm}t\in[0,t_{\mathrm{d}}]\\
(x_0-\delta x_0)\cos^{2}\left[\frac{\pi (T_2-t)}{T_2-t_{\mathrm{d}}}g(t)\right]+
\delta x_0, 
\hspace{0.1cm} t\in(t_{\mathrm{d}},T_2]
\end{array}
\right.\nonumber\\
\end{eqnarray}
where both $g(t)$ and $\lambda(t)$ are only defined in the time interval $(t_{\mathrm{d}},T_2]$, but the expression remains 
the one given in Eq.~(\ref{eq:gt}). Such a choice for $d_{-1}(t)$ ensures that it is always bounded by $\delta x_0$ and $x_0$. 
The control pulse $d_1(t)$ is then the time inverse of $d_{-1}(t)$, that is, $d_1(t)=d_{-1}(T_2-t)$. 

As outlined previously, here the goal is not only the minimization of the overlap infidelity 
$\mathcal{I}=1-\vert\langle\psi_{\mathrm{g}}\vert\psi(T_2)\rangle\vert^2$, but also the minimization of the occupancy in the middle trap. 
To this aim, we use the following objective functional\tunnu{, namely the cost function we want to minimize}:
\begin{eqnarray}
\mathcal{J}&=&1+\frac{w_E}{T_2}\int_0^{T_2}\mathrm{d}t\left[\Delta E(t)+\vert\psi_{\mathrm{d}}(x_C,t)\vert^2\right]\nonumber\\
&&-\left[\frac{w_{\mathrm{d}}}{T_2-2t_{\mathrm{d}}}\int_{t_{\mathrm{d}}}^{T_2-t_{\mathrm{d}}}\mathrm{d}t
\vert\langle\psi_{\mathrm{d}}(t)\vert\psi(t)\rangle\vert^2\right.\nonumber\\
&&+\left.w_{\mathrm{g}}\vert\langle\psi_{\mathrm{g}}\vert\psi(T_2)\rangle\vert^2\vphantom{\int}\right],
\label{eq:cost1}
\end{eqnarray}
where $\ket{\psi_{\mathrm{d}}}$ is the spatial dark state, which corresponds to the first excited state obtained by diagonalizing at each time 
the single particle Hamiltonian $\hat H(t) = \hat p^2/2m + V(x,t)$, $\Delta E(t)=\vert E_2(t)+E_0(t)-2E_1(t)\vert$ with 
$E_n(t)$ $n=0,1,2$ the first three eigenvalues at time $t$ of $\hat H(t)$ ($E_1$ is the energy of the spatial dark state), and $x_C$ 
is the position of the minimum of the middle well, where the node of the spatial dark state should be located. The energy difference 
$\Delta E(t)$ is used to keep the energy of the spatial dark state equidistant from the energies of the ground and second 
excited states, and reduce the transitions out of the dark state. The second line of Eq.~(\ref{eq:cost1}) corresponds to the 
projection of the evolved state on the actual spatial dark state in the interval $[t_{\mathrm{d}},T_2-t_{\mathrm{d}}]$, whereas the weights 
$w_E$, $w_{\mathrm{g}}$, and $w_{\mathrm{d}}$ can be adjusted for convergence. We note that a similar objective functional 
has been used in Ref.~\cite{Gajdacz2011}.

\begin{figure}[t]
\begin{center}
\includegraphics{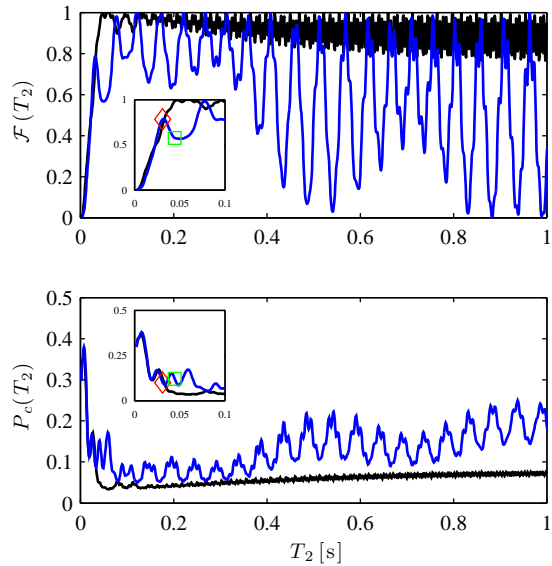}
\end{center}
\caption{(Color online). Overlap fidelity $\mathcal{F}=\vert\langle\psi_{\mathrm{g}}\vert\psi(T_2)\rangle\vert^2$ (top) and 
probability of occupancy of the middle trap (bottom) vs. time. The blue (bright) lines refer to time-dependent trap parameters $v_{\pm 1}(t)$, 
whereas the black (dark) lines refer to time-independent trap parameters $v_{\pm 1}(t)\equiv 1$. The insets show zooms of both the 
overlap infidelity (top) and the occupancy probability (bottom). The diamond and square symbols on the top of the blue (bright) line 
of both panels represent the situations for which the CRAB optimization has been performed. In both cases $\delta x_0 = 1.43\,\mu$m.}
\label{fig:IPvst}
\end{figure}

We first investigate the behavior of the SAP process without optimization. In Fig.~\ref{fig:IPvst} we show the overlap fidelity 
$\mathcal{F}=1-\mathcal{I}$ as a function of the transport time $T_2$ when the trap parameters $v_{\pm 1}(t)$ are chosen to be 
time-dependent (blue-bright line), which fixes the minima of the triple well potential to the same energy level. Instead, the 
black-dark line corresponds to the scenario for which $v_{\pm 1}(t)\equiv 1\,\forall t\in[0,T_2]$. 
In this case the first three lowest eigenstates of the Hamiltonian $\hat H(t)$ are not in resonance and the evolved state tries to follow 
the second excited eigenstate, as also illustrated in Fig.~\ref{fig:IPvst} by the behavior of the overlap infidelity at long times. 
This phenomenon occurs because when the two outer wells approach the middle one, the minima of the outer wells correspond to a larger 
energy than the minimum of the middle well (see also the magenta dashed line in Fig.~\ref{fig:potT2}), and therefore 
it is energetically more favorable for the system to follow the second excited state. 
Additionally, we note that fixing the minima of the triple well configuration to the same energy level yields a spatial dark state whose node 
is not localized in the centre of the middle trap, as desired, but it is lifted towards the outer well with lower barrier (or potential 
maximum). In order to have a node in the middle trap one should also require maxima at the same energy level, but this in not 
contemplated in our study as outlined in Sec.~\ref{sec:dipoletrap}.  Thus, the \tunnu{dark state has to be properly engineered} 
and this also explains our choice for the cost functional~(\ref{eq:cost1}).

As shown in Fig.~\ref{fig:IPvst} (bright-blue line on the top), the asymmetry of the potential, due to the fact that the maxima of the triple-well 
potential are not fixed to the same energy level (while the minima are), does not enable the atom to follow an eigenstate of $\hat H(t)$, in 
particular the spatial dark state, and an oscillatory behavior occurs. Interestingly, the occupancy in the middle trap, defined as
\begin{eqnarray}
\mathcal{P}_c(t)&=&\int_{x_L^{\max}(t)}^{x_R^{\max}(t)}\mathrm{d}x\vert\psi(x,t)\vert^2,\\
P_c(T_2)&=&\frac{1}{T_2}\int_0^{T_2}\mathrm{d}t\mathcal{P}_c(t),
\end{eqnarray}
is higher when we force the minima of the trapping potential to be at the same energy level (i.e., time-dependent $v_{\pm 1}$). 
Here $x_{L,R}^{\max}(t)$ are the positions at time $t$ of the maxima of the trapping potential of Fig.~\ref{fig:pot} (lower panel).

\begin{figure*}[htb]
\centerline{
\includegraphics[width=180mm,angle=0]{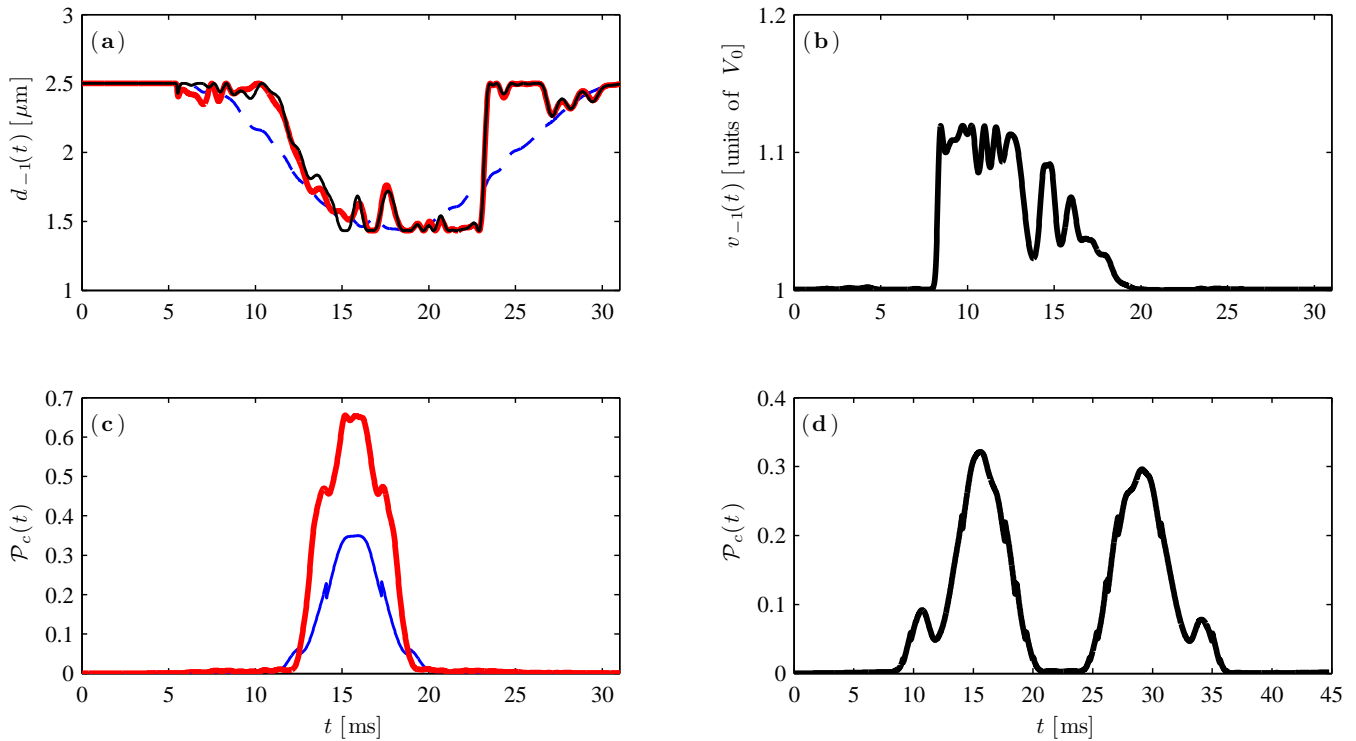}
}
\caption{(Color online). (a) Optimal control pulses obtained by means of the CRAB algorithm respectively for the three situations 
listed in the first three rows of Table~\ref{tab:2}: $T_2=$31.4$^{\star}$ms (blue dashed line), 
$T_2=$31.4$^{\dagger}$ms (black solid line), and $T_2=$31.4 ms (red solid thick line). (b) Trap parameter $v_{-1}(t)$ for the optimal 
control pulse $T_2=$31.4 ms (red solid thick line) of panel (a). (c) Probability of occupancy of the middle well vs. time obtained with 
the optimal control pulses for $T_2=$31.4$^{\star}$ms (blue thin line) and $T_2=$31.4 ms (red thick line). (d) Probability of occupancy 
of the middle well vs. time obtained with the optimal control pulse for $T_2=$44.8 ms. Here the time delay is $t_{\mathrm{d}}\simeq 5.2$ ms 
for $T_2 = 31.4$ ms, $t_{\mathrm{d}}\simeq 7.5$ ms for $T_2 = 44.8$ ms, and for both transport times $\delta x_0=1.43\,\mu$m.}
\label{fig:optdL}
\end{figure*}

In figure~\ref{fig:optdL} we show the optimal control pulses $d_{-1}(t)$ [panel (a)] together with the corresponding probabilities of occupancy 
$\mathcal{P}_c(t)$ [panel (c)] for the transport time $T_2 = 31.4$ ms, that is, the red diamond symbol in the insets of Fig.~\ref{fig:IPvst} (we 
recall that $d_1$ is the time inverted control pulse of $d_{-1}$). Instead, in panel (b) the trap parameter $v_{-1}(t)$ is displayed, which corresponds 
to the optimal quantum dynamics for $T_2 = 31.4$ ms with time-dependent trap parameters and with objective functional given in 
Eq.~(\ref{eq:cost1})  ($v_1$ is basically the time inverted of $v_{-1}$). For the CRAB optimization we used $N_g=25$ and we set $w_{\mathrm{g,d}}=0.5$ 
and $w_E=1$. Such a choice is a good trade-off between small overlap infidelity and occupancy. In Table~\ref{tab:2}, we show the results of 
both the overlap infidelity and probability of occupancy, for which a further improvement was not possible. Indeed, our attempts at optimizing 
by considering $d_{\pm 1}(t)$ as independent control pulses, by looking for an optimal set of frequencies $\omega_k$, instead of coefficients 
$\{A_k,B_k\}$, or by varying $t_{\mathrm{d}}$, have not been able to further improve the obtained results. 

Given our findings in Fig.~\ref{fig:IPvst}, we have also analyzed the scenario for which $v_{\pm 1}=1$. In this case we used for the optimization the 
following objective functional 
\begin{eqnarray}
\mathcal{J}&=& w_{\mathcal{I}}\left[1 - \vert\langle\psi_{\mathrm{g}}\vert\psi(T_2)\rangle\vert^2\right] + w_PP_c(T_2),
\label{eq:cost2}
\end{eqnarray}
where the weights $ w_{\mathcal{I}},\,w_P$ are adjusted for convergence. The results of such analysis are illustrated in both Table~\ref{tab:2} 
and Fig.~\ref{fig:optdL}.

Concerning the final probability of occupancy in the middle trap, $P_c(T_2)$, the optimization decreases the value from 0.1166, obtained with 
the initial guess control pulses $d_{\pm 1}^0$, to the value of 0.0466, obtained with the optimal control pulse, when the system tries to follow the second 
excited eigenstate; from 0.1498 to 0.0771 or to 0.0916 in the other two cases when the system is forced to follow the spatial dark state (see also Table~\ref{tab:2}). 
As illustrated in Fig.~\ref{fig:optdL} (c), the probability distribution function $\mathcal{P}_c(t)$ is peaked around $t=T_2/2$, that is, when the trap separation 
is minimum. For comparison, we also show the probability distribution for $T_2=44.8$ ms (the green square symbol in the insets of Fig.~\ref{fig:IPvst}), 
after optimization. In this situation, the distribution is almost symmetric with respect to $t=T_2/2$, because the optimized dynamics tends to split 
the transport of the wave packet from the left to the right well not directly, but in a two-step-like process, where at time $t=T_2/2$ the state is almost a dark 
state. This fact is also confirmed by the pair $(p_{\mathrm{d}},n_{\mathrm{d}})$ of values of the projection onto the instantaneous spatial dark state 
[second line of Eq.~(\ref{eq:cost1})] and the node of the spatial dark state [second integrand in the first line of Eq.~(\ref{eq:cost1})]: for $T_2=31.4$ ms 
we have (0.211,0.027), whereas for $T_2=44.8$ ms we get (0.151,0.011). Thus, for longer times, the system follows more closely the spatial dark state 
when the objective functional (\ref{eq:cost1}) is chosen. 

A final remark on the choice of the objective functionals comes from the following fact. As shown in the second and third rows of Table~\ref{tab:2}, 
both the overlap infidelity 
and the probability of occupancy in the middle well are decreased when the objective functional  (\ref{eq:cost2}) is used in the CRAB optimization. This choice 
implies a better achievement of our goals as well as a computationally less demanding simulation. Indeed, with (\ref{eq:cost2}) there is no need to diagonalize 
the instantaneous single particle Hamiltonian $\hat H(t)$. Hence, with this choice it is possible to efficiently optimize the transport in optical superlattices containing different interacting atomic species~\cite{Gajdacz2011} and, particularly relevant for our purposes, the transport of a condensate. In this case the determination of 
the instantaneous spatial dark state and its eigenvalue are more involved~\cite{Graefe2006}.  

\begin{table}
\begin{tabular}{ccccccccccccccccccccccccccccccccccccc}
\hline
\hline
\rm{$T_2$(ms)}  & & & $\mathcal{I}$ & & & $P_c(T_2)$\\
\hline
\hline
31.4$^{\star}$ & & & 0.0007 & & & 0.0466\\ 
\hline
31.4$^{\dagger}$ & & & 0.0035 & & & 0.0771\\ 
\hline
31.4 & & & 0.0048 & & & 0.0916\\ 
\hline
44.8 & & & 0.0028 & & & 0.0699 \\
\hline
\hline
\end{tabular}
\caption{\label{tab:2} Optimized errors for different transport times of the SAP technique for a single atom. The first row 
corresponds to the case for which $v_{\pm 1}=1$, whereas in the other cases $v_{\pm 1}$ are time-dependent. For the 
first two rows the objective functional used in the CRAB optimization is defined in Eq.~(\ref{eq:cost2}), \tunnu{but for the 
second row we fixed the minima of the trapping potential to the same energy level. Instead, for the last two rows, the 
objective functional is defined in Eq.~(\ref{eq:cost1})}.}
\end{table}

\begin{figure*}[t]
\begin{center}
\includegraphics[width=180mm,angle=0]{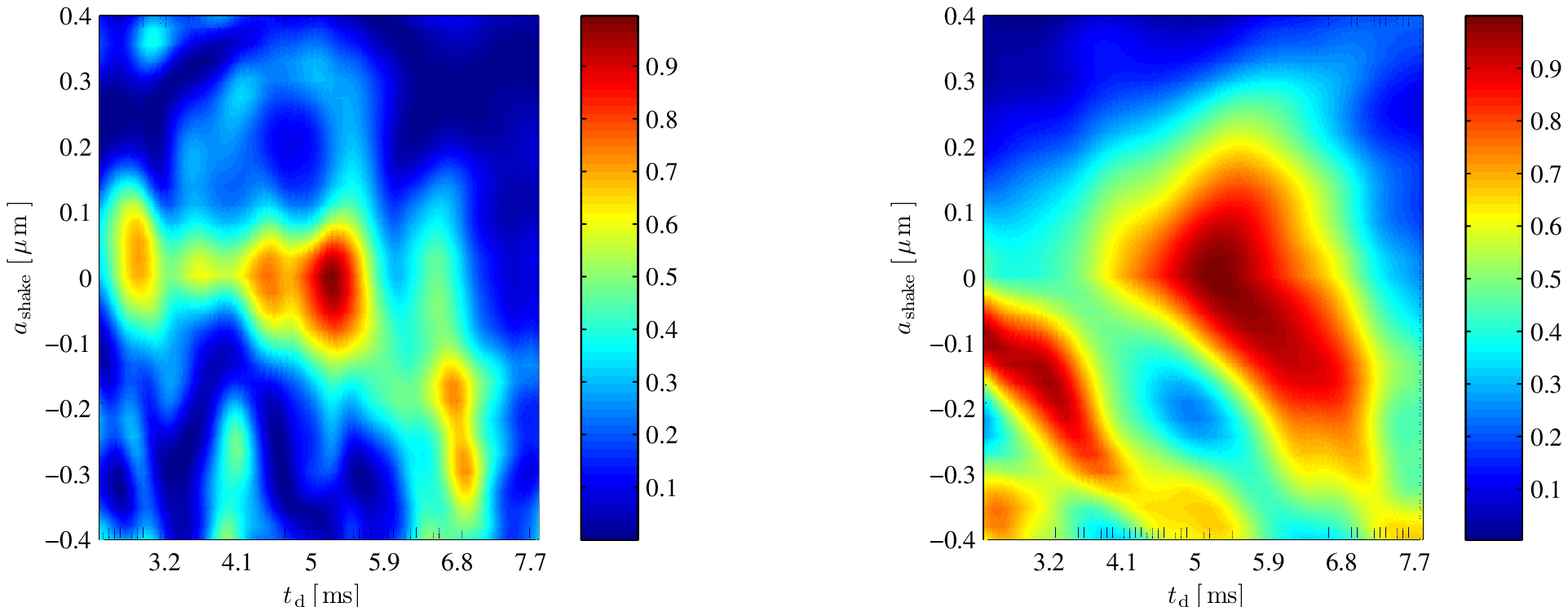}
\end{center}
\caption{(Color online). Transport efficiency (i.e., $1-\mathcal{I}$) from $\phi_{-1}^{(2)}(x)$ to $\phi_1^{(2)}(x)$ as a function of the time 
delay $t_{\mathrm{d}}$ between the two trap approaches and the amplitude $a_{\mathrm{shake}}$ of shaking the positions of 
the outer traps with $\omega_{\mathrm{shake}}=10^{-2}\omega_x$: (left) for time-dependent trap parameters $v_{\pm 1}$, that is, when 
the system is forced to follow the dark state; (right) for time-independent trap parameters $v_{\pm 1}(t)\equiv 1$, that is, when 
the system is forced to follow the second excited state. Here $\delta x_0=1.43\,\mu$m and the ``optimal" time delay  $t_{\mathrm{d}}\simeq 5.2$ 
ms.}
\label{fig:shake}
\end{figure*}

Finally, as in Sec.~\ref{sec:sp1}, we investigated the robustness of the optimized dynamics against trap position fluctuations 
by adding a shaking term to the control pulses $d_{\mp 1}^{\mathsf{opt}}(t)$, as in Eq.~(\ref{eq:shake}). Moreover, since the 
CRAB algorithm determines the optimal time-independent coefficients $A_k$ and $B_k$, we were also able 
to investigate the effect of an imprecise control of the time delay $t_{\mathrm{d}}$ on the overlap infidelity. The result of such analysis is 
illustrated in Fig.~\ref{fig:shake} for $\omega_{\mathrm{shake}}/\omega_x=10^{-2}$ and for $T_2=31.4$ ms. We compare the obtained 
results for two cases: when the atomic system is forced to follow the instantaneous spatial dark state (right) and when it is forced to follow 
the second excited state of the trapping potential (left). As already pointed out in Ref.~\cite{Eckert2004}, for purely harmonic confinement, 
the SAP process is less affected by imprecise timing, but the overlap infidelity drops faster for fluctuations in the trap positions than in the 
ideal case considered in Ref.~\cite{Eckert2004}. Indeed, while for harmonic traps the fidelity is reduced by 1-2\%~ \cite{Eckert2004}, 
at the optimal time delay and $a_{\mathrm{shake}}\sim\ell_x/2$ ($\ell_x$ is the harmonic oscillator ground 
state width), for Gaussian traps  the fidelity $\lesssim$ 60\% (right picture), which shows how detrimental the anharmonicity of the trapping 
potential is~\cite{QIP:Negretti05}. However, this phenomenon is also due to the fact that the system follows the second excited state, which 
is more sensitive to energy losses, and therefore to a worsening of the transport fidelity. Additionally, the figure shows the fragility of the 
dynamics that the instantaneous spatial dark state of the system is forced to follow, especially concerning the control of the time delay. This further 
analysis confirms what we already noticed in the overlap infidelity and occupancy probability, that is, it is more efficient to follow the second 
excited state rather than the spatial dark one.

%%

%%---- Section --------------------------------------------------------------------------------------

%%

\section{Transport of a condensate}

In this section we investigate the optimal transport of a BEC in optical dipole potentials such as the ones in Eq.~(\ref{eq:Vxt}).
We assume the quasi-1D regime of quantum degeneracy and a mean field description of the atomic system dynamics, that is, 
we assume that the Gross--Pitaevskii equation 
(GPE)~\cite{Pitaevskii2003} 
\begin{align}
-i\hbar\frac{\partial\psi(x,t)}{\partial t}&=\hat H_{\mathrm{gp}}[\psi]\psi(x,t),\nonumber\\
\phantom{=}\hat H_{\mathrm{gp}}[\psi]&=\left[-\frac{\hbar^2}{2m}\frac{\partial^2}{\partial x^2}+V(x,t)+g_{1D}N\vert\psi(x,t)\vert^2\right]
\label{eq:Hgp}
\end{align}
well describes the physics of our problem. Here $\psi$ is normalized to one, $N $ is the atom number, 
$g_{1D}=2\hbar\omega_{\perp}a_{3D}^{\mathrm{s}}\left(1-1.4603\frac{a_{3D}^{\mathrm{s}}}{a_{\perp}}\right)^{-1}$~\cite{Olshanii1998}, 
$a_{3D}^{\mathrm{s}}$ is the three dimensional (3D) s-wave scattering length, and $a_{\perp}=(2\hbar/m\omega_{\perp})^{1/2}$. This 
assumption implies that the radial confinement is frozen to its 
ground state, and therefore that the ratio $\eta=\omega_{\perp}/\omega_x$ is significantly larger than 1 ($\omega_{\perp}$ is the 
transverse trap frequency, that is, the trap frequency in the $y-z$ plane). As we already pointed out in 
Ref.~\cite{QIP:Treutlein06b}, a good value is $\eta=20$, for which radial excitations due to two-body collisions can be suppressed. 
We underscore that a simulation of the current experimental setup would require a 3D simulation of the GPE, since the 
transport occurs in the transverse direction while the axial confinement has a shallower trap than the transverse one, where the transport 
process occurs. 

%%

%%---- Section --------------------------------------------------------------------------------------

%%

\subsection{Attractive inter-particle interaction}
\begin{figure}[t]
\begin{center}
\includegraphics{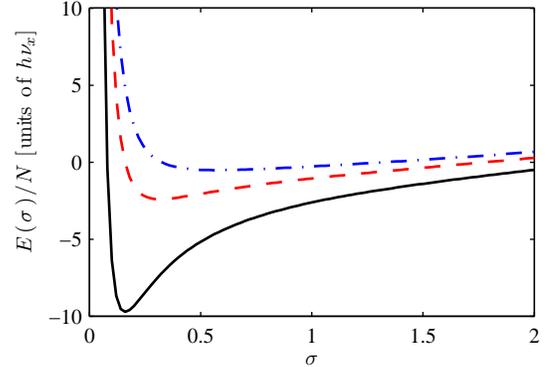}
\end{center}
\caption{(Color online).  Gross-Pitaevskii energy functional vs. the effective width in the Gaussian model for attractive interacting atoms in the 
quasi 1D regime. The black (solid) line corresponds to $\eta=20$ and $N=10$, the red (dashed) line to $\eta=20$ and $N=5$, and the blue (dashdot) line 
to $\eta=5$ and $N=10$. For all curves we considered $^{85}$Rb atoms confined in a harmonic trap with frequency $\omega_x = 2\pi \times 711$ Hz.}
\label{fig:Ncrit}
\end{figure}

It is well known (see, for instance, Ref.~\cite{Pitaevskii2003}), that for attractive interactions (i.e., $g_{1D}<0$, that is, $a_{3D}^{\mathrm{s}}<0$), a critical number of condensed atoms exists, $N_{\mathrm{cr}}$, such that for $N>N_{\mathrm{cr}}$ the condensate is not stable and the GPE has no longer a stationary solution. We have investigated this phenomenon in the quasi 1D regime by considering the Gross-Pitaevskii energy functional. 
For an arbitrary confinement potential $V(x)$ the functional is defined as~\cite{Pitaevskii2003}:
\begin{align}
\label{eq:gpef}
\frac{E}{N}&=\!\int\!\mathrm{d}x\left[
\frac{\hbar^2}{2m}\left\vert\frac{\partial\psi(x)}{\partial x}\right\vert^2
+V(x)\vert\psi(x)\vert^2%\right.\nonumber\\
%\phantom{=}&\left.
+g_{1D}\frac{N}{2}\vert\psi(x)\vert^4
\right].
\end{align}
For a harmonic trap, a Gaussian Ansatz for the condensate wave function can be used, which has been proven to provide an excellent estimation of 
$N_{\mathrm{cr}}$ for a three-dimensional BEC~\cite{Pitaevskii2003}. To this aim, let us consider the following Ansatz for the condensate wave function 
(normalized to 1)
\begin{eqnarray}
\label{eq:psiGauss}
\psi(x)=(\sigma\ell_x\sqrt{\pi})^{-1/2}\exp\left(-\frac{x^2}{2\sigma^2\ell_x^2}\right),
\end{eqnarray}
where $\sigma$ is a dimensionless parameter which represents the effective width of the BEC. By inserting (\ref{eq:psiGauss}) into (\ref{eq:gpef}) we 
obtain
\begin{align}
\label{eq:gpefho}
\frac{E}{N\hbar\omega_x}&=
\frac{1}{4}\left(\frac{1}{\sigma^2}+\sigma^2\right)-\frac{f_{\mathrm{c}}\alpha}{\sigma\sqrt{2\pi}},
\end{align}
where $f_{\mathrm{c}}^{-1}=1-1.4603 a_{3D}^{\mathrm{s}}/a_{\perp}$, and $\alpha=\eta N a_{3D}^{\mathrm{s}}/\ell_x$. 
The behavior of the GP energy functional, for some values of $\eta$ and $N$ for an atomic cloud of $^{85}$Rb atoms 
trapped in the hyperfine state $\ket{F=2,m_F=-2}$, is illustrated in Fig.~\ref{fig:Ncrit}. As the figure shows, the local minimum 
disappears for either a small atom number (solid vs. dashed lines) 
or for a small ratio $\eta$ (solid vs. dashdot lines). Importantly, the energy local minimum appears always for $E<0$, that is, 
the interaction energy, $E_{\mathrm{int}}$, exceeds the 
kinetic energy. Contrarily to the repulsive case, for attractive interatomic forces the kinetic energy, $E_{\mathrm{kin}}$, cannot 
be neglected. Indeed,  it stabilizes the condensate against collapses, namely the condensate is stable as long as 
$E_{\mathrm{kin}}>E_{\mathrm{int}}$. We can give a rough estimation of $N_{\mathrm{cr}}$ by using this inequality:
\begin{eqnarray}
\label{eq:kinint}
E_{\mathrm{kin}}\sim - \frac{\hbar^2}{2m\ell_x^2}
\qquad 
E_{\mathrm{int}}\sim -2\hbar\omega_{\perp}f_{\mathrm{c}}N\frac{\vert a_{3D}^{\mathrm{s}}\vert}{\ell_x},
\end{eqnarray}
which imply
\begin{eqnarray}
\label{eq:Ncrit}
N_{\mathrm{cr}}\simeq(4\eta f_{\mathrm{c}})^{-1}\frac{\ell_x}{\vert a_{3D}^{\mathrm{s}}\vert}.
\end{eqnarray}
By fixing $\eta=20$, we have $N_{\mathrm{cr}}\simeq 0.3$, for $\omega_x = 2\pi \times 711$ Hz, and $N_{\mathrm{cr}}\simeq 2.7$, 
for $\omega_x = 2\pi \times 7.11$ Hz. The latter value of $N_{\mathrm{cr}}$ could be in principle further enhanced by reducing $\eta$, 
even though the quasi 1D condition would no longer be very well fulfilled. Hence, from this analysis, we see that, for an attractive BEC in the 
quasi 1D regime, the admissible condensate atom number is extremely small. With such condensate atomic numbers the realization 
of attractive BECs in optical microtraps is actually not feasible. Given this, the attractive case will be discarded in the subsequent transport analysis.

%%

%%---- Section --------------------------------------------------------------------------------------

%%

\subsection{Optimization of step 1 of the transport process}

Firstly, we computed the ground state of the left well by using the imaginary time propagation technique. We start by considering $3 N$ 
atoms in such a way that we have precisely $N$ atoms in each well. This approach is valid when the three wells are far apart. The imaginary time propagation 
for $3N$ atoms yields a wave function which is the sum of three inverted parabolas, for large $N$, or almost Gaussian functions, for small $N$. 
Each of these three spatially separated density profiles is localized in one of the three wells. Then, we selected the part of the wave function localized 
in the left well, properly normalized, as initial state. This new wave function describes precisely a BEC with $N$ atoms [we have checked the 
stationarity of the solution by propagating it in real time via GPE with $N$ in the mean-field potential and as confinement the initial (static) 
potential]. When the two outer traps move towards the middle one, since the Gaussian potentials do overlap, the three wells cannot be 
treated as independent anymore (see also Fig.~\ref{fig:pot} lower panel). Even though the condensate wave functions of each of the three wells do not 
overlap, the curvature of the outer wells is different from the middle one. Hence, it is no longer straightforward to determine how many atoms 
are contained in each well. To overcome this problem, we propagate adiabatically the condensate wave function trapped in the well centered 
in $x_0^{\prime}$ towards the one centered in $x_0$. The trap position $x_0$ is chosen as the minimum separation between the three wells such that 
$N\vert\psi(x,T_1)\vert^2$ is localized only on the left well, that is, the atomic density in the middle and right wells can be effectively neglected. 
We underscore that the value of $x_0$ crucially depends on $N$, for fixed $g_{1D}$. Hence, the adiabatically evolved condensate wave function 
is chosen as goal state of step 1 of the transport process.  We then analyzed how fast the initial state can be propagated towards the goal state, 
by simulating the dynamics for different transport times $T_1$, which have been chosen much smaller than the adiabatic transfer time.

We have considered atomic ensembles with $N=10$ or 200 $^{87}$Rb atoms per well (the latter have been recently obtained in 
experiments~\cite{Lengwenus2010}). For 10 atoms the initial potential configuration is the same as for the single particle scenario 
(Fig.~\ref{fig:pot} top), with $x_0=-3.0\,\mu$m and a slightly smaller trap frequency $\omega_x=2\pi\times 702$ Hz with respect to $^{85}$Rb. 
For $N=200$, since the condensate wave function has a much larger width than the single particle one, the dipole potential has to be adjusted. To 
this aim, in order to keep the (initial) lattice periodicity fixed, that is, $\vert x_0^{\prime}\vert = 6.5\,\mu$m (i.e., with laser beam waist 1.3 $\mu$m), the 
potential depth has been increased up to $V_0=k_B\times 25\,\mu$K. Such a potential depth implies a single-well trap frequency 
$\omega_x= (4V_0/m\sigma^2)^{1/2} \simeq 2\pi\,12$ kHz, very similar to the trap geometry of Ref.~\cite{Lengwenus2010}, and $x_0=-3.8\,\mu$m as 
minimal (target) separation.

As initial guess for the control pulse we used
\begin{eqnarray}
\!\!\!\!\!\!\!\!  D_{\mathsf{ho}}(t)=\left\{
\begin{array}{lllll}
\frac{\upsilon_m^2t^2}{\Delta x} & & & & 0\le t <\frac{\Delta x}{2\upsilon_m},\\
\upsilon_m t - \frac{\Delta x}{4} & & & & \frac{\Delta x}{2\upsilon_m}\le t < \frac{\Delta x}{\upsilon_m}\\
\frac{\upsilon_m^2(T_1-t)^2}{2(\Delta x-\upsilon_mT_1)}+\Delta x & & & & \frac{\Delta x}{\upsilon_m}\le t \le T_1
\end{array}
\right.,
\label{eq:muga}
\end{eqnarray}
where $\Delta x = \vert x_0^{\prime}\vert -\vert x_0\vert$, and $\upsilon_m=3\Delta x /2T_1$ is the maximum trap velocity during the transport. 
Such a control pulse has been proven to be optimal for a 1D condensate in a moving harmonic potential at the transport times 
$T_{1,n}=3(2n+1)\pi/\omega_x$ with $n\in\mathbb{N}$~\cite{Torrontegui2011}. Thus, there exists a minimum transport time, 
$T_{1,0}=3\pi/\omega_x$, for which no excitation in the condensate is produced. 

\begin{figure}[t]
\begin{center}
\includegraphics{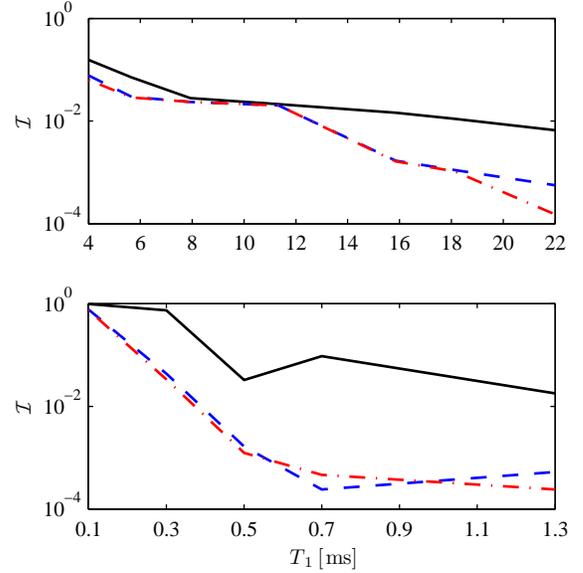}
\end{center}
\caption{(Color online).  Overlap infidelity vs. transport time: black (solid) line with the control pulse $D_{\mathsf{ho}}(t)$ 
defined in Eq.~(\ref{eq:muga}); the other two lines show the infidelity obtained with the CRAB optimized control pulse 
$d_{\mathsf{opt}}(t)=D_{\mathsf{ho}}(t)g_{\mathsf{opt}}(t)$ for $N_g=8$ (dashed line) and $N_g=16$ (dashdot line). The 
upper panel corresponds to $N=10$, whereas the lower one to $N=200$. }
\label{fig:tab2}
\end{figure}

In exactly the same way as for the single-atom transport, we investigated the (quantum) speed limit of step 1 of the transport process, 
whose results are illustrated in Fig.~\ref{fig:tab2} for $N=10$ (top) and $N=200$ (bottom) atoms with repulsive interaction. For 
200 atoms the potential depth is about $V_0\simeq 43.5 \hbar \omega_x$, the chemical potential $\mu=39.2 \hbar\omega_x$ whereas 
in the Thomas-Fermi limit we have $\mu_{\mathrm{TF}}=37.4\hbar\omega_x$~\cite{Molmer1998}. Thus, the system is well within this limit. 
Concerning the optimization, the CRAB algorithm works precisely as we described in Sec.~\ref{sec:sp1}, with the only difference that we 
have to substitute the Schr\"odinger equation with the GPE and define the control pulse as $d(t)=D_{\mathsf{ho}}(t)g(t)$, where $g(t)$ is 
given by Eq.~(\ref{eq:gt}). Besides, the overlap infidelity is defined again through Eq.~(\ref{eq:infidelity}), where 
$\psi_{\mathrm g}^{\prime}(x)\equiv\psi(x,T_{\mathrm{ad}})$ is the state obtained adiabatically, $T_{\mathrm{ad}}\sim 3$ ms, starting 
from the ground state of the left well of the initial potential configuration with trap separation $\vert x_0^{\prime}\vert = 6.5\,\mu$m. 
The same procedure is used for $N=10$ atoms. 

We see from Fig.~\ref{fig:tab2} that, while for $N=10$ the infidelity decreases monotonically with respect to the transport time $T_1$, for 
$N=200$ this is not the case, and it becomes a monotonic function only for $N_g=16$. We attribute this behavior of the infidelity to a non 
perfect revival of Bogoliubov excitation modes present during the transport process, which have a larger impact for bigger condensates, 
because of the larger non-linear interaction. 
To further improve the results one could also optimize the dynamics of the Bogoliubov collective excitations, for instance, by solving the 
time-dependent Bogoliubov-de Gennes equations~\cite{Castin1998}. This approach, however, would allow to engineer the 
Bogoliubov modes, but at the expenses of a very demanding numerical optimization.

Furthermore, as shown in Fig.~\ref{fig:tab2}, the transport times for $N=200$ are shorter than in the single-particle and small 
condensate cases. This is basically due to a shorter transport distance $\Delta x=2.7\,\mu$m (in the single-particle case 
$\Delta x=4\,\mu$m) and to the trap frequency, which is $\sim 16$ times larger than in the single atom scenario. Furthermore, 
we see that the control pulse~(\ref{eq:muga}) is an excellent guess with satisfactory overlap infidelities up to 1 ms for 200 
atoms and up to 16 ms for 10 atoms, even for transport times $T_1\ne T_{1,n}$. Notably, with respect to the single particle 
optimization, the addition of harmonics does not improve significantly the overlap infidelity for short transport times. This behavior 
may also be related to the initial guess for the coefficients $\{A_k,B_k\}$, for which we always started by setting their initial values to zero. 
Indeed, this may occur also for $T_1=0.7$ ms ($N=200$), where the control pulse with $N_g=16$ yields a slightly worst overlap 
infidelity with respect to the one obtained with $N_g=8$. The choice for the initial values of $\{A_k,B_k\}$ might be not the right one, 
since the control landscape may have several minima: the larger the number $N_g$ is, the larger the control landscape. Thus, our 
initial choice likely produces an optimal control solution trapped in a local minimum that is not the same for a larger $N_g$. 
We also note that by performing the optimization on the frequencies $\omega_k$ instead of the coefficients $\{A_k,B_k\}$ 
the improvement in the infidelity is very small.

Regarding the quantum speed limit~\cite{Giovannetti2003}, it can be roughly fixed to 0.5 ms for $N=200$, which is larger than 
$1/\nu_x\simeq 0.08$ ms and is slightly smaller than $T_{1,0}\simeq 0.13$ ms. \tunnu{We (numerically) defined the limit by 
considering the time for which the infidelity is approximately $10^{-3}$. This is a reasonable threshold to quantify the 
error on the distance between the state evolved until time $T_1$ and the goal state. We note that the quantum speed limit is 
roughly determined by $\max_{t\in[0,T_1]}\{h/[E_g(t)-E_e(t)]\}$, where $h$ is the Planck constant, $E_g(t)$ is the instantaneous 
ground state energy, and $E_e$ is the instantaneous energy of the first excited state. For shorter times, it is not physically possible 
to bring the system in the ground state of the trap without populating excited energy levels, which, on the other hand, are needed, during 
the interval $(0,T_1)$, to perform a fast transport of the atom.} Instead, for $N=10$ atoms, the quantum speed limit can be roughly fixed to 
$T_1=16$ ms, where the infidelity is about $10^{-3}$. Even though in the single particle scenario we had a slightly higher value of the 
trap frequency, because of the use of $^{85}$Rb atoms, it is interesting to note that already a small atomic cloud significantly alters the 
quantum speed limit, which, for a single atom, has been estimated around 3 ms.

\begin{figure}[t]
\begin{center}
\includegraphics{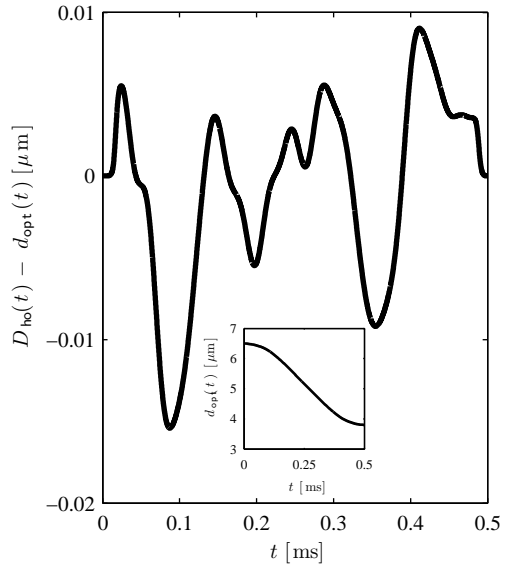}
\end{center}
\caption{(Color online). Difference between the guess control pulse $D_{\mathsf{ho}}(t)$ and the optimal one of step 1 of the 
transport process for $T_1=0.5$ ms and $N_g=16$ ($N=200$). In the inset the optimal control pulse is displayed. }
\label{fig:optdbec}
\end{figure}

\begin{figure}[t]
\begin{center}
\includegraphics{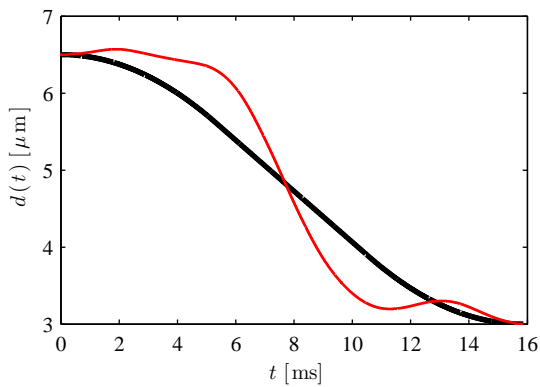}
\end{center}
\caption{(Color online). Control pulses of the step 1 of the transport process for $N=10$ atoms: initial guess given by 
Eq.~(\ref{eq:muga}) thick (black) line, and optimal CRAB pulse with $N_g=16$ thin (red) line. The transport time is $T_1\simeq 15.9$ 
ms.}
\label{fig:optdbecsmall}
\end{figure}

In Fig.~\ref{fig:optdbec} the difference between the initial guess (\ref{eq:muga}) and the optimal control pulse for $N=200$ atoms, $T_1 = 0.5$ 
ms and $N_g=16$ is depicted. This plot shows how small is the correction on the guess control pulse, even though it is quite important 
to decrease by more than an order of magnitude the overlap infidelity. For $N=10$ atoms, the initial guess pulse (thick line in 
Fig.~\ref{fig:optdbecsmall}) is rather different with respect to the CRAB optimized one. This larger distortion is due to the fact that since the 
potential depth $V_0=2.55\hbar\omega_x$ and (initial) trap separation are the same as in the single particle scenario, the potential wells are 
not deep enough to consider the control pulse of Eq.~(\ref{eq:muga}), optimal for a harmonic trap, as a good transport pulse. 
Indeed, while the single-particle energy is about 0.46$\hbar\omega_x$, the chemical potential for 10 atoms is $\mu\simeq1.87\hbar\omega_x$.

Finally, we also investigated the robustness of the optimal control pulse for $N=200$ against fluctuations of the outer trap positions like for the single 
atom dynamics. For the transport time $T_1=0.5$ ms the optimal solution obtained with CRAB is rather robust: the overlap infidelity 
changes from 0.0012 to 0.0046 for $a_{\mathrm{shake}}=\ell_x \simeq 0.1\,\mu$m. This effect is due to the cooperative behavior of the 
atoms in the collective motion of the condensate. Instead, we did not investigate the effect of dimensionality, because, unlike in the single 
particle scenario, the nonlinear term appearing in the GPE is also affected by the augmented space geometry, and therefore the comparison 
would not be fair (apart from the issue of validity in the quasi-2D regime).

%%

%%---- Section --------------------------------------------------------------------------------------

%%

\subsection{Optimization of step 2: SAP process}

The optimization of SAP with interacting particles is more difficult with respect to the single atom scenario. Indeed, as also discussed 
in Ref.~\cite{Graefe2006}, in the spectrum of the nonlinear Gross-Pitaevskii Hamiltonian~(\ref{eq:Hgp}) loops near the avoided crossing 
points and new eigenstates of $\hat H_{\mathrm{gp}}$ emerge when enhancing the nonlinear interaction. As pointed out by Graefe $et$ 
$al.$~\cite{Graefe2006}, within a three-mode model, SAP, in order to work in the nonlinear regime, has to fulfill the following two 
conditions: (i) $g_{1D}N\Delta\ge 0$; (ii) $g_{1D}N/\ell_x<g_{\mathrm{c}}=\Delta$. Here $\Delta$ represents a detuning between the three 
wells, that is, the resonance condition needed for SAP. We note that the resonance condition in this case imposes that the onsite energies 
of the wells, $\hbar\omega_k(t)+\mu_k(t)$, are constant at all times, where $\omega_k(t)$ is the local frequency of the $k$-th well and 
$\mu_k(t)$ the corresponding chemical potential at time $t$. The inequality (ii) shows that there exists an upper bound on the nonlinear interaction 
strength for the realization of SAP. The problem we are studying, however, cannot be strictly treated within a three-mode approximation. 
Nevertheless the model will be used as a guideline when discussing the results of the optimization.

\begin{figure}[t]
\begin{center}
\includegraphics{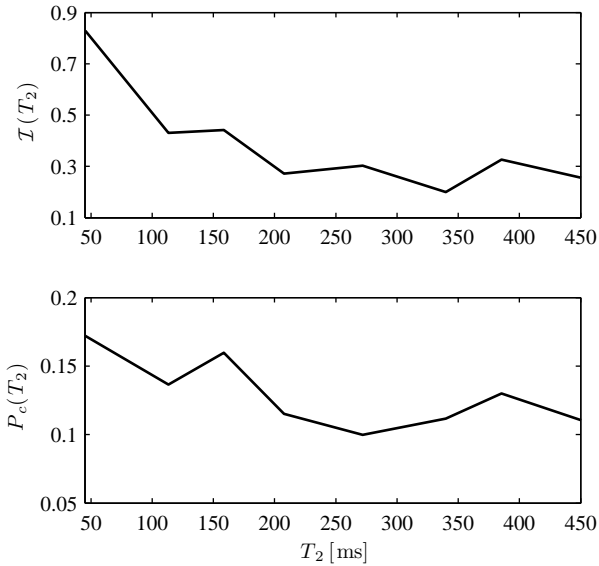}
\end{center}
\caption{(Color online). Overlap infidelity (top) and probability of occupancy of the middle trap (bottom) vs. time for the 
optimization of the SAP protocol for $N=10$ interacting $^{87}$Rb atoms.  
The minimum allowed trap separation is $\delta x_0=1.42$ $\mu$m and $v_{\pm 1}(t)=1$.}
\label{fig:sapbec}
\end{figure}

As for the single particle study, we applied the CRAB algorithm in order to understand whether optimal control can improve 
the performance of the SAP protocol. Both for $N=50$ and $200$, however, we noticed that for a fixed number of harmonics 
($N_g=10$) CRAB was not able to reduce the value of the overlap infidelity obtained with the initial guess control 
pulses~(\ref{eq:dpmguess}). This (empirical) observation holds both when we are optimizing the control pulse by searching 
for the optimal set of coefficients $A_k$, $B_k$, and when we seek the optimal set of frequencies $\omega_k$. Moreover, we 
numerically noticed that the convergence of the algorithm to the value of the overlap infidelity obtained with the initial guess control 
pulse takes longer than in the single-atom case. Even though in these two cases the number of atoms is likely much larger than 
the one allowed for the realization of SAP in the nonlinear regime, we attribute the occurrence of such a phenomenon to a more 
elaborated control landscape topology, that is, a control landscape with a large number of local minima due to the emergence 
of new eigenstates in the system. We did not further investigate this aspect, which deserves a deeper analysis in a separated 
work, but we rather chose to further reduce the number of atoms to $N=10$. In this case CRAB was able to improve the performance 
of the protocol with respect to the initial guess control pulse. As already pointed out in the previous section, with respect to the 
single-atom scenario, here we used $^{87}$Rb atoms which imply a smaller trap frequency $\omega_x$ and  a trap separation $x_0=-3.0$ 
$\mu$m. Apart from these small changes, due to a different atomic species and a broader size of the atomic sample, the trap 
configuration is essentially the one of the single-atom case. Nevertheless, the optimization carried out for different transport times 
$T_2$ could not go below $\sim$20\% of overlap infidelity and $\sim$10\% of population in the middle trap. The result of such a study 
is illustrated in Fig.~\ref{fig:sapbec}.
The obtained results cannot be improved by further optimizing the frequencies $\omega_k$. This shows that even though optimal 
control can improve the performance of the protocol, there is however a physical limit due to the SAP resonance condition for which no 
further optimized dynamics can be achieved. Indeed, at $t=T_2/2$ the separation between the wells is minimal and we can roughly 
estimate the detuning as $\Delta\simeq 0.15\,\hbar\omega_x$, whereas $g_{1D}N/\ell_x\simeq 5.56\,\hbar\omega_x$, which shows how 
condition (ii) is not satisfied even with only $N=10$ atoms. To increase $\Delta$ one should further reduce $\delta x_0$, but then the three 
wells merge in a single one, or, alternatively, by reducing the atom number. In this case, however, the BEC would be very small and the GP 
description might be also questionable. Although with a different trap setup, the analysis carried out in Ref.~\cite{Morgan2011} also 
shows that the overlap infidelity increases quite quickly with the number of atoms and that even with only two $^{87}$Rb atoms the 
(non-optimized) performance of SAP is quickly harmed ($\sim$16\% of infidelity). Besides, as Fig.~\ref{fig:sapbec} illustrates, the behavior is not 
monotonic, which is probably related to a non optimal dynamics of the Bogoliubov modes. 

Finally, concerning the population of the middle trap, Fig.~\ref{fig:sapbec} shows that it is almost constant with a minimum of about 0.1. 
We note that, in comparison with the single-atom case, we did not further minimize the population of the middle well, since the transport 
efficiency was already lower, and therefore we preferred to focus on the minimization of the overlap infidelity [i.e., we set $w_P=0$ 
in Eq.~(\ref{eq:cost2})]. Nevertheless, the CRAB  optimization has been able to further reduce the population with respect to the one 
obtained with the initial guess control pulse. 

%%

%%---- Section --------------------------------------------------------------------------------------

%%

\section{Conclusions}

In this paper we have numerically investigated the performance of the SAP protocol by means of optimal control both at the single particle 
and at the many-body level. In our analysis we have considered trap parameters, atomic species, and atom numbers that are used in 
current experiments~\cite{Schlosser2011}. The transport process has been split in three steps, because of the initial large trap 
separation. The first step brings the atom(s) localized in the left well closer to the middle well in such a way that tunneling between 
the three wells occurs, therefore enabling the realization of the second step of the transport, that is, the SAP process. Afterwards, 
the third step of the transport process brings further away from the middle well the atom(s) localized in the right well. 
We have seen that while we can easily achieve the quantum speed limit, both for the single particle and the condensate scenario, 
for the first and last steps of the dynamical transport process, the second one requires a higher degree of control already for small 
transport time reductions with respect to the ``adiabatic" times. In the single atom case, we observe a smaller population in the middle 
trap when the system is forced to follow the second excited state of the trap (i.e., time-independent $v_{\pm 1}$) rather than following 
the actual dark state (i.e., time-dependent $v_{\pm 1}$). In the latter case, due to the different energy level of the maxima of the triple 
well configuration, the node of the dark state wave function is not localized within the middle trap, but outside. This fact forced us to 
additionally engineer the shape of the dark state wave function rendering the control landscape more complicated. Thus, we had 
to make a trade-off between transfer efficiency and suppression of the middle trap population. In addition, we observed that the 
engineering of the dark state reduces the robustness against trap and time delay fluctuations of the optimal control pulse. We note 
that, in order to further improve the transfer efficiency and reduce the population of the middle trap, by engineering properly 
the potential, one could employ a programmable and computer controllable nematic liquid-crystal spatial light modulator, where the 
trap separation can be varied by changing the periodicity of the modulator~\cite{QIP:Bergamini04}. Alternatively, optical superlattices 
can be used, which would allow to fix the three minima at the same energy level as well as the two maxima. 

The optimization of the SAP protocol for a condensate strongly relies on the atom number and onsite energy of the wells. We have 
investigated in some detail the performance of the protocol for $N=10$ atoms with repulsive interaction. The analysis showed that the 
CRAB algorithm is able to improve the transport efficiency with respect to the one obtained with the initial guess control pulse, but the 
maximum attainable efficiency, for a transport time not longer than 450 ms, is about 80\% with a population in the middle well of about 
10\%. It is not clear whether longer times could yield a better efficiency, which would require a longer computational time, but if this 
would be the case, one has also to take into account the effects of decoherence. For instance, if we consider atom chip 
technology~\cite{QIP:ACbook11}, where the expected limits due to surface-induced decoherence of motional states are comparable to 
the ones of the hyperfine states, which have coherence times of about 1s~\cite{Treutlein2004}, our analysis already shows that 
we are actually close to the limit of the SAP protocol. This ultimate limit, for a relatively small BEC, is due to the emergence of 
new eigenstates and crossing levels, as discussed in detail in Ref.~\cite{Graefe2006}, which break down the SAP protocol when the 
nonlinear interaction exceeds a critical value. 

In summary, from our investigations, it emerges that while at the single atom level SAP can be optimized below the 0.1\% level, and 
possibly observed in current experiments, the application of an optimized SAP technique to a condensate is rather limited, already 
even with small number of atoms. On the other hand, it would be interesting to investigate more precisely and more generally the 
influence of the nonlinear interaction of BEC on the quantum speed limit of a certain dynamical process, and this will be pursued in 
future work.

%%

%%---- Acknowledgments -----------------------------------------------------------------------------------

%%

\section*{Acknowledgments}

A.N. is grateful for the invitation to Universitat Aut\`{o}noma de Barcelona and thanks Tommaso Caneva for useful hints in the 
implementation of the CRAB algorithm. We acknowledge financial support from the EU Integrated Project AQUTE, QIBEC, PICC,
the Deutsche Forschungsgemeinschaft within the Grant No. SFB/TRR21 (T.C.),
the Marie Curie Intra European Fellowship  (Proposal No. 236073, OPTIQUOS) within the 7th European Community Framework 
Programme (A.N.), financial support through Spanish MICINN contracts FIS2008-02425 and CSD2006-00019, the Catalan 
Government contract SGR2009-00347, and Grant No. AP 2008-01275 from the Spanish MICINN FPU Program (A. B.).

\section*{Appendix}

The determination at each time of $v_{-1}(t)$ and $v_1(t)$ is a rather complicated nonlinear minimization problem. 
In our simulations, however, we noticed that an excellent approximation to the values of $v_{-1}(t)$ and $v_1(t)$ 
is given by the following procedure: at the beginning the positions of the minima of the trapping potential 
(\ref{eq:Vxt}) are determined by looking for the roots $\{x_L,x_C,x_R\}$ of the function
\begin{equation}
V^{\prime}(x,t)=  \sum_{k=-1}^1 [x-k d_k(t)]\exp\left\{
-\frac{(x-k d_k(t))^2}{2 w^2}
\right\},
\end{equation}
where $v_{\pm 1}=1$. Then, we use the following formulae:
\begin{widetext}
\begin{eqnarray}
v_{-1}(t) &=& 
\left\{\left[\exp\left(-\frac{x_C^2}{2w^2}\right)-\exp\left(-\frac{x_L^2}{2w^2}\right)\right] 
    \left[\exp\left(-\frac{(x_C-d_1(t))^2}{2w^2}\right)-\exp\left(-\frac{(x_R-d_1(t))^2}{2w^2}\right)\right]\right.\nonumber\\
    &-&\left.\left[\exp\left(-\frac{x_C^2}{2w^2}\right)-\exp\left(-\frac{x_R^2}{2w^2}\right)\right]
    \left[\exp\left(-\frac{(x_C-d_1(t))^2}{2w^2}\right)-\exp\left(-\frac{(x_L-d_1(t))^2}{2w^2}\right)\right]\right\}\nonumber\\
    &/&\left\{\left[\exp\left(-\frac{(x_C-d_1(t))^2}{2w^2}\right)-\exp\left(-\frac{(x_L-d_1(t))^2}{2w^2}\right)\right]
    \left[\exp\left(-\frac{(x_C+d_{-1}(t))^2}{2w^2}\right)-\exp\left(-\frac{(x_R+d_{-1}(t))^2}{2w^2}\right)\right]\right.\nonumber\\
    &-&\left.\left[\exp\left(-\frac{(x_C+d_{-1}(t))^2}{2w^2}\right)-\exp\left(-\frac{(x_L+d_{-1}(t))^2}{2w^2}\right)\right] 
    \left[\exp\left(-\frac{(x_C-d_1(t))^2}{2w^2}\right)-\exp\left(-\frac{(x_R-d_1(t))^2}{2w^2}\right)\right]\right\},\nonumber\\
\end{eqnarray}    
\begin{eqnarray}    
v_1(t) &=& \frac{v_{-1}(t)\left[\exp\left(-\frac{(x_C+d_{-1}(t))^2}{2w^2}\right)-\exp\left(-\frac{(x_L+d_{-1}(t))^2}{2w^2}\right)\right]
                   -\exp\left(-\frac{x_C^2}{2w^2}\right)-\exp\left(-\frac{x_L^2}{2w^2}\right)} 
      {\exp\left(-\frac{(x_C-d_1(t))^2}{2w^2}\right)-\exp\left(-\frac{(x_L-d_1(t))^2}{2w^2}\right)} 
%\left\{\left[\exp\left(-\frac{x_C^2}{2w^2}\right)-\exp\left(-\frac{x_L^2}{2w^2}\right)\right] 
%    \left[\exp\left(-\frac{(x_C-d_R(t))^2}{2w^2}\right)-\exp\left(-\frac{(x_R-d_R(t))^2}{2w^2}\right)\right]\right.\nonumber\\
%    &-&\left.\left[\exp\left(-\frac{x_C^2}{2w^2}\right)-\exp\left(-\frac{x_R^2}{2w^2}\right)\right]
%    \left[\exp\left(-\frac{(x_C-d_R(t))^2}{2w^2}\right)-\exp\left(-\frac{(x_L-d_R(t))^2}{2w^2}\right)\right]\right\}\nonumber\\
 %   &/&\left\{\left[\exp\left(-\frac{(x_C-d_R(t))^2}{2w^2}\right)-\exp\left(-\frac{(x_L-d_R(t))^2}{2w^2}\right)\right]
 %   \left[\exp\left(-\frac{(x_C+d_L(t))^2}{2w^2}\right)-\exp\left(-\frac{(x_R+d_L(t))^2}{2w^2}\right)\right]\right.\nonumber\\
 %   &-&\left.\left[\exp\left(-\frac{(x_C+d_L(t))^2}{2w^2}\right)-\exp\left(-\frac{(x_L+d_L(t))^2}{2w^2}\right)\right] 
 %   \left[\exp\left(-\frac{(x_C-d_R(t))^2}{2w^2}\right)-\exp\left(-\frac{(x_R-d_R(t))^2}{2w^2}\right)\right]\right\}\nonumber\\ 
 %   &\cdot&\frac{\exp\left(-\frac{(x_C+d_L(t))^2}{2w^2}\right)-\exp\left(-\frac{(x_L+d_L(t))^2}{2w^2}\right)} 
 %     {\exp\left(-\frac{(x_C-d_R(t))^2}{2w^2}\right)-\exp\left(-\frac{(x_L-d_R(t))^2}{2w^2}\right)} 
 %     -\frac{\exp\left(-\frac{x_C^2}{2w^2}\right)-\exp\left(-\frac{x_L^2}{2w^2}\right)}
 %     {\exp\left(-\frac{(x_C-d_R(t))^2}{2w^2}\right)-\exp\left(-\frac{(x_L-d_R(t))^2}{2w^2}\right)}.  
\end{eqnarray}
\end{widetext}
These solutions are obtained by solving the system of linear equations: $V(x_L,t)=V(x_C,t),\,V(x_R,t)=V(x_C,t)$. 

Finally, we also mention that our numerical simulations of both the Schr\"odinger and the Gross-Pitaevskii equation have been 
performed by means of the split operator technique together with the fast Fourier transform algorithm~\cite{Press2007}.

%%

%%---- Section --------------------------------------------------------------------------------------

%%

%\bibliographystyle{atchip}
\bibliography{letteratura,atomchips}

\end{document}